\documentstyle[preprint,aps]{revtex} 
\begin{document}
\draft 
\tightenlines
\title{Chiral Sum Rules to Second Order in Quark Mass}
\author{Eugene Golowich} 
\address{Department of Physics and Astronomy,
University of Massachusetts \\
Amherst MA 01003 USA}
\author{Joachim Kambor}
\address{Institut f\"ur Theoretische Physik,
Universit\"at Z\"urich \\
CH-8057 Z\"urich, Switzerland}
\maketitle
\begin{abstract}
\noindent
A new calculation of the isospin and hypercharge
axialvector current propagators ($\Delta_{{\rm A}33}^{\mu\nu}(q^2)$
and $\Delta_{{\rm A}88}^{\mu\nu}(q^2)$) to two loops in 
$SU(3)\times SU(3)$ chiral perturbation theory is used 
to derive chiral spectral function sum rules valid to 
second order in the light quark masses.  Explicit forms are given 
for the three-pion isospin axialvector spectral functions at low 
energy and application of the sum rules to the determination 
of counterterms of the chiral lagrangian is discussed.  

\end{abstract}

Chiral sum rules were introduced some thirty years ago 
by Weinberg, who used chiral symmetry to derive two sum rules 
valid in the limit of massless light quarks.~\cite{sw1}   
The derivation of several other sum rules soon followed, 
likewise valid only in the chiral limit.~\cite{{dgmly},{dmo}}      
These efforts predated the rise of quantum 
chromodynamics (QCD) as a theory of the strong interactions.  
A subsequent QCD derivation of the Weinberg sum rules revealed 
that the first could be extended to the real world of nonzero 
$u$, $d$, $s$ quark mass, but not the second.~\cite{fnr}  
Due to the analytic intractability of low energy QCD, 
theoretical understanding of the chiral sum rules 
away from the chiral limit has been limited.  
In this Letter we show how chiral perturbation 
theory (ChPT)~\cite{{sw},{gl}} can be used to obtain a number of 
chiral sum rules, mostly new, to second order in quark mass.  

The derivation is made possible by a recent calculation of 
axialvector current propagators through two-loop order in 
ChPT.~\cite{gk}  This leads to both the derivation of spectral 
function sum rules for nonzero quark mass 
and the prediction of axialvector spectral functions at low energy. 
It also suggests an application of the sum rules to evaluation 
of ${\cal O}(p^6)$ counterterms which appear as part of 
the ChPT procedure.~\cite{fs}  

The $SU(3)$ axialvector current propagators are defined as 
\begin{equation}
\Delta_{{\rm A}ab}^{\mu\nu}(q^2) \equiv i \int d^4x~ e^{iq\cdot x}~
\langle 0|T\left( A^\mu_a (x) A^\nu_b (0)\right)|0\rangle 
\qquad (a,b = 1,\ldots, 8)
\label{aa1}
\end{equation}
and have spectral content 
\begin{equation}
{1\over \pi} {\cal I}m~\Delta_{{\rm A}ab}^{\mu\nu}(q^2) = 
(q^\mu q^\nu - q^2g^{\mu\nu}) \rho_{{\rm A}ab}^{(1)}
(q^2 ) + q^\mu q^\nu \rho_{{\rm A}ab}^{(0)} (q^2) \ , 
\label{aa2}
\end{equation}
where $\rho_{{\rm A}ab}^{(1)}$ and $\rho_{{\rm A}ab}^{(0)}$ are 
the spin-one and spin-zero spectral functions.  The tensor structure 
of Eq.~(\ref{aa2}) motivates the decomposition usually adopted in 
the literature, 
\begin{equation}
\Delta_{{\rm A}ab}^{\mu\nu}(q^2) = 
(q^\mu q^\nu - q^2 g^{\mu\nu}) \Pi_{{\rm A}ab}^{(1)}
(q^2 ) + q^\mu q^\nu \Pi_{{\rm A}ab}^{(0)} (q^2) \ \ ,
\label{aa2a}
\end{equation}
where $\Pi_{{\rm A}ab}^{(1)}$ and $\Pi_{{\rm A}ab}^{(0)}$ are 
respectively the spin-one and spin-zero axialvector polarization 
functions.  As regards flavor notation, we consider $a=b=3$ (isospin), 
$a=b=8$ (hypercharge) and understand throughout that $aa = 33,88$ is 
not summed.  Determination of the 
axialvector current propagators in ChPT~\cite{stern} proceeds 
by introducing axialvector sources into the chiral lagrangians 
and calculating all amplitudes which connect single source 
external states.  The procedure is basically as 
described in Ref.~\cite{gk1}.  

We summarize briefly the results through two-loop order, 
suppressing flavor labelling for simplicity.  For the 
tree-level propagator, one finds 
\begin{equation}
\Delta_{{\rm A},\mu\nu}^{\rm (tree)}(q^2) = F^2_0~ g_{\mu\nu} - 
{F^2_0 \over q^2 - m^2}~q_\mu q_\nu \ \ ,
\label{aa3}
\end{equation}
where $m$ and $F_0$ are the mass and decay constant parameters 
to leading order in the chiral expansion.  The proper behavior 
is seen to occur in the chiral limit $m \to 0$, 
where the propagator takes on a purely spin-one 
(`transverse') form consistent with current conservation 
$\partial_\mu A^\mu = 0$.  Rewriting the above 
in terms of the tensor structure of Eq.~(\ref{aa2}) yields 
\begin{equation}
\Delta_{{\rm A},\mu\nu}^{\rm (tree)}(q^2) 
= - F^2_0 \left[ {1\over q^2}~(q_\mu q_\nu - q^2 g_{\mu\nu}) 
+ {m^2\over q^2(q^2 - m^2)}~q_\mu q_\nu \right] \ \ .
\label{tree}
\end{equation}
There are kinematic poles at $q^2 = 0$ in both the spin-one 
and spin-zero polarization functions, although the sum 
$\Pi_{\rm A}^{(1)}$ + $\Pi_{\rm A}^{(0)}$ is free of such poles.  

We have determined the propagator through two-loop order 
({\it i.e.} summing over tree, one-loop and two-loop 
contributions) to have the structure 
\begin{equation}
\Delta_{{\rm A},\mu\nu}(q^2) =
(F^2 + {\hat \Pi}_{\rm A}^{(0)}(q^2)) 
g_{\mu\nu}  - {F^2 \over q^2 - M^2} q_\mu q_\nu 
+ ( 2L_{10}^{\rm r} - 4H_1^{\rm r} + {\hat \Pi}_{\rm A}^{(1)}(q^2))
 (q_\mu q_\nu - q^2 g_{\mu\nu}) \ ,
\label{two-loop}
\end{equation}
where $F^2$, $M^2$ are now renormalized at two-loop level,  
$L_{10}^{\rm r}$, $H_1^{\rm r}$ are ${\cal O}(p^4)$ counterterms 
which appear in the one-loop analysis and 
${\hat \Pi}_{\rm A}^{(0,1)}(q^2)$ are finite two-loop 
functions.\footnote{The quantity $H_1^{\rm r}$ is regularization 
dependent, as are analogous terms in ${\hat \Pi}^{(0)}(0)$. 
However, such unphysical contributions are absent from 
the spectral function sum rules.}     
The process of deriving Eq.~(\ref{two-loop}), describing the 
renormalization procedure and displaying the various formulae is quite
lengthy and will be deferred to another setting.~\cite{gk}  However, 
${\hat \Pi}_{\rm A}^{(0)}(q^2)$ can be shown to vanish 
in the limit of zero quark mass and thus the above expression has 
the proper chiral behavior.  The relation between the amplitudes 
appearing in Eq.~(\ref{aa2a}) with those in Eq.~(\ref{two-loop}) is 
\begin{eqnarray}
\Pi_{{\rm A}}^{(1)}(q^2) &=& 
2L_{10}^{\rm r} - 4H_1^{\rm r} + {\hat \Pi}_{\rm A}^{(1)}(q^2)
- {F^2 + {\hat \Pi}_{{\rm A}}^{(0)}(q^2) \over q^2} \ , \nonumber \\
\Pi_{{\rm A}}^{(0)}(q^2) &=& {{\hat \Pi}_{{\rm A}}^{(0)}(q^2) \over q^2} 
- {F^2 M^2 \over q^2 (q^2 - M^2)} \ \ .
\label{relate}
\end{eqnarray}
We now turn to some implications of the two-loop analysis.  

In chiral perturbation theory, the axialvector spectral 
functions $\rho_{{\rm A}ab}^{(0,1)}$ 
first receive contributions at chiral order $p^6$.   
Their threshold behavior can be deduced directly from Eq.~(\ref{aa2}) 
via the imaginary parts of the two-loop propagators.  Such 
contributions arise entirely 
from the set of (`sunset') diagrams having three-particle intermediate 
states.  Figures~1,2 display the predicted low energy behavior 
of the three-pion contributions to the isospin spectral functions 
$\rho_{{\rm A}33}^{(1)}[3\pi]$ and $\rho_{{\rm A}33}^{(0)}[3\pi]$.  
The spectral function $\rho_{{\rm A}33}^{(1)}[3\pi]$ has been 
inferred phenomenologically from $3\pi$ emission in tau decay 
over much of the allowed energy range,~\cite{aleph} although more 
data near threshold is required to decisively test the ChPT
prediction.  To our knowledge no phenomenological determination 
of $\rho_{{\rm A}33}^{(0)}[3\pi]$ yet exists, and we will address 
this topic in a future work.  

An independent procedure for calculating the spectral functions is to 
employ unitarity by inserting all possible three-pion intermediate 
states in the two-point functions.~\cite{unit},\cite{cfu}  
We find the two-loop and unitarity schemes to yield precisely 
the same results. This serves as a highly nontrivial check on the 
correctness of the two-loop functions ${\hat \Pi}_{\rm A}^{(0,1)}(q^2)$ as 
well as on the constituitive relations in Eq.~(\ref{relate}). 

The derivation of chiral sum rules proceeds by first obtaining 
dispersion theoretic expressions for the various polarization 
functions.  The asymptotic behavior ($s \to \infty$) of the 
vector and axialvector spectral functions which follows from 
QCD is given by~\cite{fnr},\cite{bnp} 
\begin{eqnarray}
& & \rho_{{\rm V}aa}^{(1)}(s) \sim {\cal O}(1) \ , \quad 
\rho_{{\rm A}aa}^{(1)}(s) \sim {\cal O}(1) \ , \quad 
\rho_{{\rm A}aa}^{(0)}(s) \sim {\cal O}(s^{-1}) \ ,
\quad (\rho_{{\rm A}33}^{(1)} - \rho_{{\rm A}88}^{(1)})(s) \sim 
{\cal O}(s^{-1}) \ , 
\nonumber \\ 
& & (\rho_{{\rm V}aa}^{(1)} - \rho_{{\rm A}aa}^{(1)})(s) \sim 
{\cal O}(s^{-1}) \ , \qquad 
(\rho_{{\rm V}aa}^{(1)} - \rho_{{\rm A}aa}^{(1)} 
- \rho_{{\rm A}aa}^{(0)})(s) \sim {\cal O}(s^{-2})\ \ ,
\label{asym}
\end{eqnarray}
where $\rho_{{\rm V}aa}^{(1)}$ is defined analogously to 
$\rho_{{\rm A}aa}^{(1)}$ ({\it cf}~Eq.~(\ref{aa2})) and 
$\rho_{{\rm V}aa}^{(0)}$ vanishes since we assume 
isospin symmetry.  
The information in Eq.~(\ref{asym}) can be used, together with 
analyticity and the corresponding asymptotic behavior of the 
polarization functions, to derive 
dispersion relations for the vector and axialvector polarization 
functions.  The dispersion relations occur in three classes, 
beginning with those which combine vector and axialvector 
amplitudes, {\it e.g.} 
\begin{equation}
\left(\Pi_{{\rm V}aa}^{(1)} - \Pi_{{\rm A}aa}^{(1)}
- \Pi_{{\rm A}aa}^{(0)}\right)(q^2) = \int_0^\infty ds ~ 
{(\rho_{{\rm V}aa}^{(1)} - \rho_{{\rm A}aa}^{(1)} 
- \rho_{{\rm A}aa}^{(0)})(s) \over s - q^2 - i \epsilon} 
\ \ . \label{aa4} 
\end{equation}
Due to the highly convergent behavior of $(\rho_{{\rm V}aa}^{(1)} 
- \rho_{{\rm A}aa}^{(1)} - \rho_{{\rm A}aa}^{(0)})(s)$ at large $s$, 
this combination of polarization functions is `superconvergent', {\it i.e.} 
the first moment $q^2~\left(\Pi_{{\rm V}aa}^{(1)} - \Pi_{{\rm A}aa}^{(1)}
- \Pi_{{\rm A}aa}^{(0)}\right)(q^2)$ also obeys an unsubtracted 
dispersion relation.  Then there are dispersion relations for the 
axialvector polarization functions of a given flavor, {\it e.g.} 
\begin{equation}
q^2 \Pi_{{\rm A}aa}^{(0)}(q^2) - \lim_{q^2 = 0}\left( q^2 
\Pi_{{\rm A}aa}^{(0)}(q^2)\right) = q^2 \int_0^\infty ds 
{\rho_{{\rm A}aa}^{(0)}(s) \over s - q^2 - i \epsilon} 
\qquad 
\ , \label{aa5} 
\end{equation}
where we work with $q^2 \Pi_{{\rm A}aa}^{(0)}(q^2)$ due to 
the presence of $q^2 = 0$ kinematic poles.  The quantity  
$q^2 \Pi_{{\rm A}aa}^{(1)}(q^2)$ obeys an analogous relation.  
Finally, there are dispersion relations 
for $SU(3)$-breaking combinations, such as 
$(\Pi_{{\rm A}33}^{(1)} + \Pi_{{\rm A}33}^{(0)}
- \Pi_{{\rm A}88}^{(1)} - \Pi_{{\rm A}88}^{(0)})(q^2)$  and 
$q^2~ (\Pi_{{\rm A}33}^{(1)} - \Pi_{{\rm A}88}^{(1)})(q^2)$.  

Sum rules are obtained by evaluating arbitrary derivatives of such 
relations at $q^2 = 0$.  We consider first the two most well-known 
examples.  Thus, evaluation at $q^2 = 0$ of the dispersion relation for 
$q^2~\left(\Pi_{{\rm V}aa}^{(1)} - \Pi_{{\rm A}aa}^{(1)}
- \Pi_{{\rm A}aa}^{(0)}\right)(q^2)$ yields Weinberg's first 
sum rule,~\cite{sw1}    
\begin{equation}
F_a^2 = \int_0^\infty ds~ 
(\rho_{{\rm V}aa}^{(1)} - \rho_{{\rm A}aa}^{(1)} 
- {\bar \rho}_{{\rm A}aa}^{(0)})(s) \ \ ,
\label{aa9}
\end{equation}
where we have defined ${\bar \rho}_{{\rm A}aa}^{(0)}(s) \equiv 
\rho_{{\rm A}aa}^{(0)}(s) - F_a^2 \delta(s - M_a^2)$.  Note that 
the sum rule is now evaluated {\it away} from the chiral limit.   
Likewise Eq.~(\ref{aa4}) at $q^2 = 0$ yields the inverse-moment 
sum rule~\cite{{dmo},{etc}} 
\begin{equation}
\left(\Pi_{{\rm V}aa}^{(1)} - \Pi_{{\rm A}aa}^{(1)}
- \Pi_{{\rm A}aa}^{(0)}\right)(0) = \int_0^\infty ds ~ 
{(\rho_{{\rm V}aa}^{(1)} - \rho_{{\rm A}aa}^{(1)} 
- \rho_{{\rm A}aa}^{(0)})(s) \over s}\ \ , 
\label{aa10}
\end{equation}
which now includes two-loop contributions on the left-hand-side (LHS).  

More generally, entire sequences of sum rules are derivable, such as 
those for a given flavor, 
\begin{eqnarray}
& & {1\over n!}\left[{d \over dq^2}\right]^n\left( 
\Pi_{{\rm V}aa}^{(1)} - \Pi_{{\rm A}aa}^{(1)}
- \Pi_{{\rm A}aa}^{(0)}\right)(0) = \int_0^\infty ds ~ 
{(\rho_{{\rm V}aa}^{(1)} - \rho_{{\rm A}aa}^{(1)} 
- \rho_{{\rm A}aa}^{(0)})(s) \over s^{n+1}} \quad (n \ge 0)\ , 
\label{aa11} \\
& & {1\over n!}\left[{d \over dq^2}\right]^n 
{\hat \Pi}_{{\rm A}aa}^{(0)}(0) = \int_0^\infty ds ~ 
{{\bar \rho}_{{\rm A}aa}^{(0)}(s) \over s^n} \qquad (n \ge 1) \ ,
\label{aa11a} \\
& & {1\over (n -1)!}\left[{d \over dq^2}\right]^{n-1} 
{\hat \Pi}_{{\rm A}aa}^{(1)}(0) - {1\over n!}\left[{d \over dq^2}\right]^n 
{\hat \Pi}_{{\rm A}aa}^{(0)}(0) = \int_0^\infty ds ~ 
{\rho_{{\rm A}aa}^{(1)}(s) \over s^n} \qquad (n \ge 2) \ \ .
\label{aa11b} 
\end{eqnarray}
A sequence of sum rules explicitly involving broken $SU(3)$ is 
\begin{equation}
{1\over n!}\left[{d \over dq^2}\right]^n\left( 
\Pi_{{\rm A}33}^{(1)} + \Pi_{{\rm A}33}^{(0)}
- \Pi_{{\rm A}88}^{(1)} - \Pi_{{\rm A}88}^{(0)}\right)(0) 
= \int_0^\infty ds ~ {(\rho_{{\rm A}33}^{(1)} + \rho_{{\rm A}33}^{(0)} 
- \rho_{{\rm A}88}^{(1)} - \rho_{{\rm A}88}^{(0)})(s)  \over s^{n+1}} \ ,
\label{aa12} 
\end{equation}
with $n \ge 0$.  

Various applications of chiral sum rules have appeared in the 
literature since their introduction. 
These range from obtaining relations between 
the spectra of vector and axialvector mesons~\cite{sw1}
to applying the full battery of data and theory inputs 
to phenomenologically test the Weinberg sum rules 
in their zero mass setting.~\cite{dg94}  Here we show 
how to obtain numerical estimates for certain finite 
${\cal O}(p^6)$ counterterms associated with the ChPT 
renormalization procedure.~\cite{gk2}  Recall that in a two-loop 
calculation the leading divergences must be 
cancelled by ${\cal O}(p^6)$ counterterms $\{ B_k \}$.  
Each such counterterm can be expressed in dimensional regularization as 
\begin{equation}
B_k = \mu^{2(d - 4)} \sum_{n=2}^{-\infty} \ B_k^{(n)}(\mu) 
~{\overline\lambda}^n 
= \mu^{2(d - 4)} \left[~ B_k^{(2)}(\mu) ~{\overline\lambda}^2 
\ + \ B_k^{(1)}(\mu) ~{\overline\lambda} \ + \ B_k^{(0)}(\mu) \ 
+ \dots ~\right] 
\label{aa14}
\end{equation}
where ${\overline \lambda}$ is the singular quantity 
\begin{equation}
{\overline \lambda} \equiv 
{ 1 \over 16 \pi^2} \left[ {1 \over d - 4} - {1\over 2}
\left( \log {4\pi} - \gamma + 1 \right) \right] \ \ .
\label{aa15}
\end{equation}
The $\{ B_k^{(2)}(\mu)\}$ and $\{ B_k^{(1)}(\mu)\}$ are chosen 
to subtract off all divergences, and a large number of these 
so-called $\beta$-functions is thereby determined.~\cite{gk} 
However, the physical result will contain the finite quantities 
$\{ B_k^{(0)}(\mu)\}$.\footnote{The dependence on renormalization scale 
$\mu$ is known explicitly from renormalization group equations.} 
Like the renormalized electron charge and mass 
in QED, these finite counterterms must somehow 
be determined from experiment.  Thus, for example 
consider the sum rule of Eq.~(\ref{aa11b}) with $n=2$ and isospin 
flavor,   
\begin{equation}
4 \left( 2 B_{32}^{(0)} - B_{33}^{(0)}\right)(\mu)  
- {1\over 3072 \pi^4 F_\pi^2} \left( 0.204 + \log{M_\pi^2 \over \mu^2} 
+ {5\over 4} \log{M_K^2 \over \mu^2} \right) 
= \int_0^\infty ds ~ 
{\rho_{{\rm A}33}^{(1)}(s) \over s^2} \ \ ,
\label{aa13} 
\end{equation}
where the first term inside the parentheses on the LHS 
arises from a scale-independent two-loop contribution.  From the 
renormalization procedure, one knows already that 
$\left( 2 B_{32}^{(2)} - B_{33}^{(2)}\right)(\mu) = 0$ 
and  $\left( 2 B_{32}^{(1)} - B_{33}^{(1)}\right)(\mu) = 
3/1024 \pi^2 F_\pi^2$.  To estimate 
$\left( 2 B_{32}^{(0)} - B_{33}^{(0)}\right)(\mu)$, 
we approximate the spectral function in terms of 
the $a_1$ resonance contribution taken in narrow width approximation, 
$\rho_{{\rm A}33}^{(1)}(s) \simeq g_{a_1} \delta (s - M^2_{a_1})$.  
From the fit of Ref.~\cite{dg94} and adopting the
renormalization scale $\mu = M_{a_1}$, we obtain 
$(2 B_{32}^{(0)} - B_{33}^{(0)})(M_{a_1}) \simeq 0.0030~{\rm GeV}^{-2}$.  
The ${\cal O}(p^6)$ counterterm dominates the other terms on the 
LHS of Eq.~(\ref{aa13}), showing the importance of a 
full ChPT calculation as compared to a chiral-log treatment.  
A more thorough phenomenological analysis will involve use of the 
entire spectrum and include error bars in the final estimate.  
However, this simple example serves to demonstrate the general procedure.  

\vspace{0.2cm}

To conclude, a new two-loop ChPT calculation of axialvector 
current propagators, as embodied by Eq.~(\ref{two-loop}), has 
been performed.  In addition to yielding a complete two-loop 
renormalization of the pion and eta masses and decay constants, it 
has led to predictions for axialvector spectral functions and to the 
derivation of spectral function sum rules.  An application of 
the sum rules to determine finite ${\cal O}(p^6)$ counterterms 
has been provided.  Additional work will involve careful analysis of 
the existing database to provide as precise a determination of 
the counterterms as experimental uncertainties allow as well 
as addressing the phenomenological extraction of spectral functions 
like $\rho_{{\rm A}33}^{(0)}[3\pi]$.  

We thank John Donoghue, J\"urg Gasser and Marc Knecht for 
useful comments.  This work was supported by the US National 
Science Foundation and by Schweizerischer Nationalfonds.

\newpage

\begin{center}
{\large\bf FIGURE CAPTIONS}
\end{center}

\noindent 1. Three-pion contribution to spin-one isospin spectral function. 

\noindent 2. Three-pion contribution to spin-zero isospin spectral function.

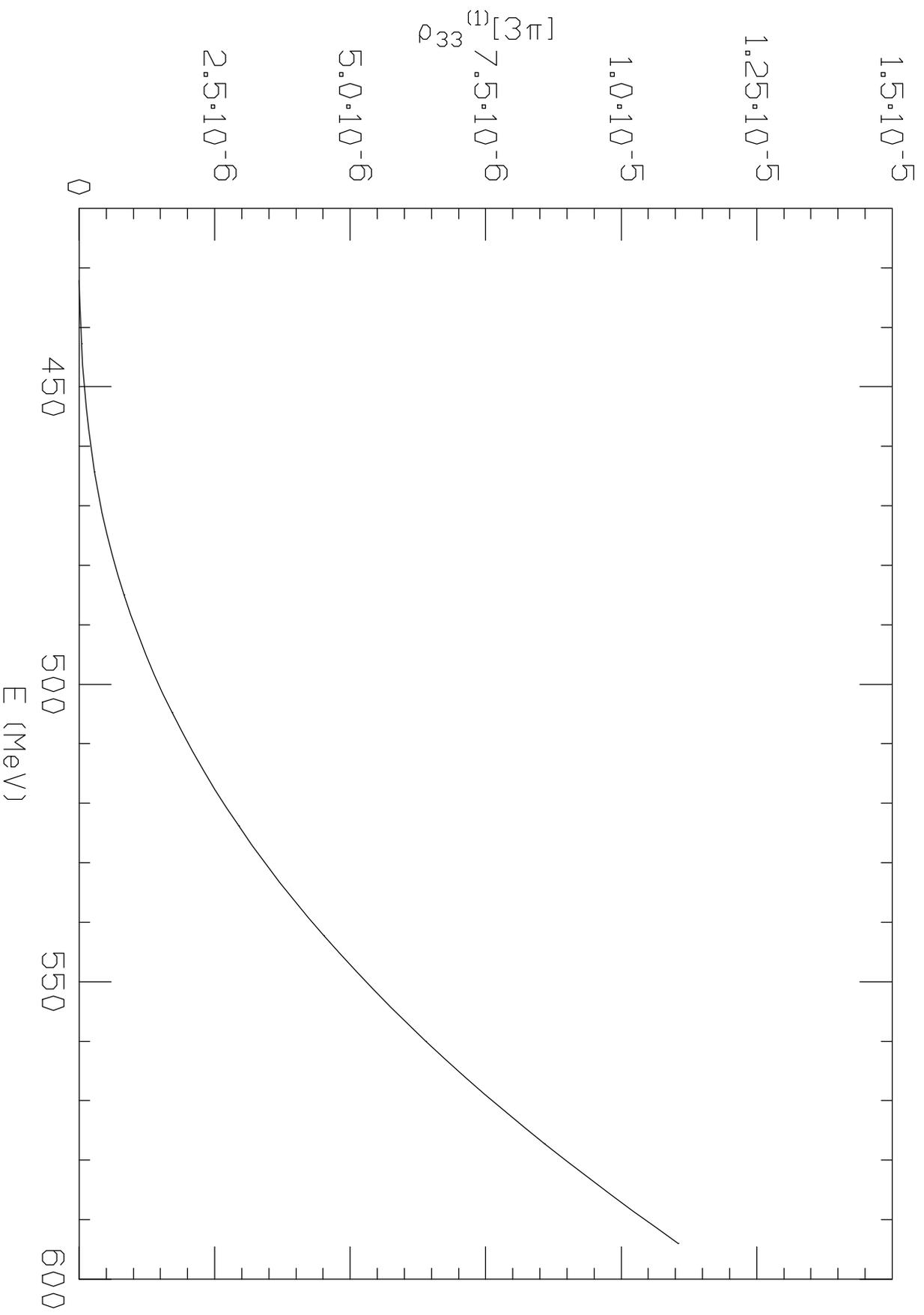
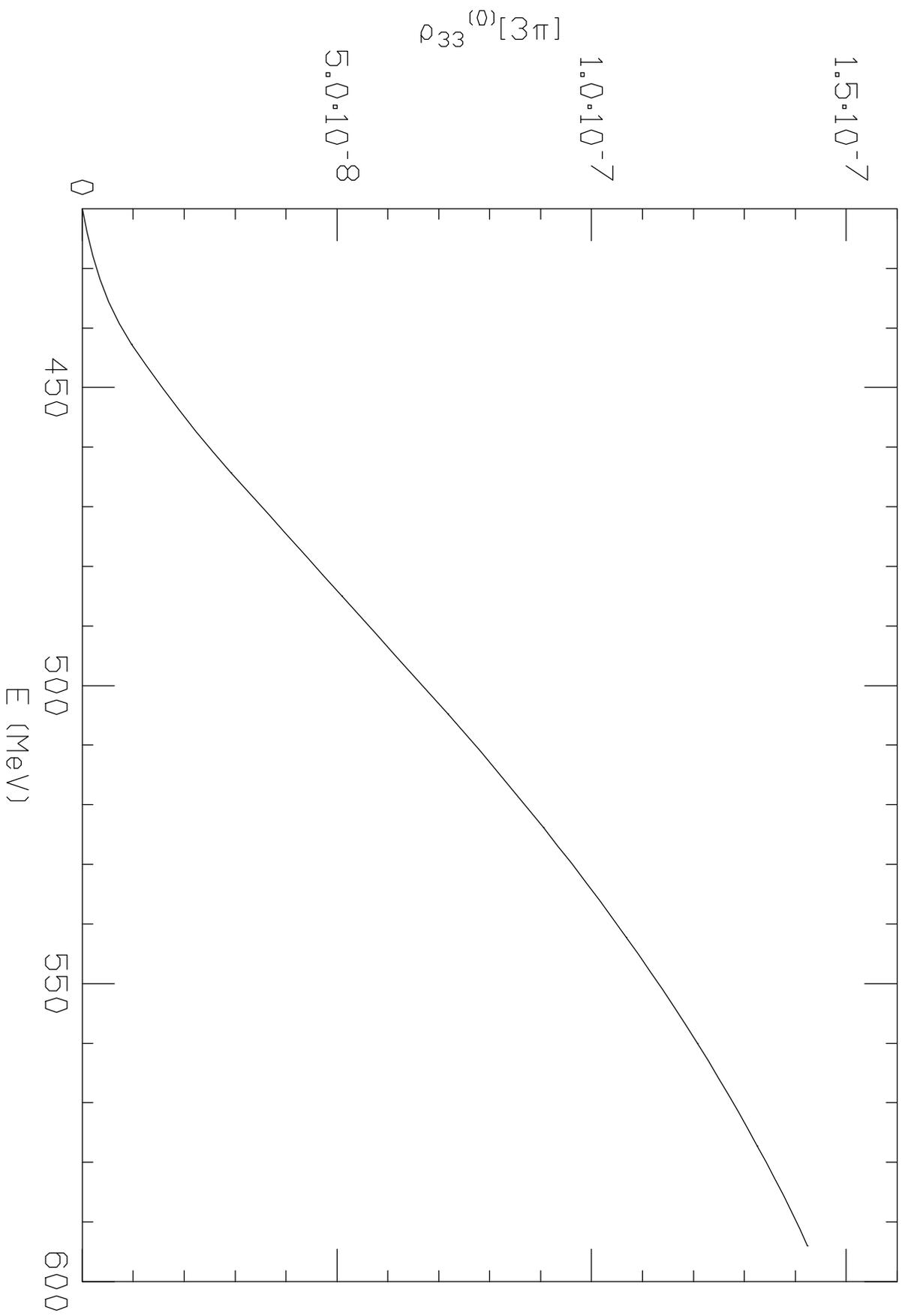
\end{document}